\documentclass[12pt]{article}
\usepackage{amsmath,amssymb,ccaption,amsmath,graphicx,harvard,mathdots, comment}
 \usepackage{threeparttable, booktabs, multirow} 
  \usepackage{setspace}
  \usepackage{hyperref} 
\usepackage{array}
\newcolumntype{C}{>{\centering\arraybackslash}p{1.1cm}}
 
\begin{document}

\title{Forecasting for monetary policy}

\author{Laura Coroneo\thanks{University of York, Department of Economics and Related Studies, Email: laura.coroneo@york.ac.uk.
\newline
The author thanks the editor Pierre Pinson and two anonymous referees for their constructive and insightful comments, which helped improve the paper substantially. I also thank David Hendry, Paulo Santos Monteiro and Tim Willems, and participants to the Durham-York-Newcastle workshop for comments on the preliminary draft.}}
\date{2 April 2025}
\maketitle

\begin{abstract}
\noindent This paper discusses three key themes in forecasting for monetary policy highlighted in the \citeasnoun{bernanke2024forecasting} review: the challenges in economic forecasting, the conditional nature of central bank forecasts, and the importance of forecast evaluation. In addition,  a formal evaluation of the Bank of England's inflation forecasts indicates that, despite the large forecast errors in recent years, they were still accurate relative to common benchmarks.
\end{abstract}



\clearpage
\begin{doublespace}
\section{Introduction}\label{sec_intro}
In the wake of the largest inflation surge witnessed by major developed countries in over two decades, the role of central banks in managing inflation risks has never been more scrutinized.

The recent \citeasnoun{bernanke2024forecasting} review offers a fresh lens on the Bank of England’s forecasting practices, providing recommendations on how central bank forecasts should be approached and utilised.  The review clarifies that central banks rely on economic forecasts to assist in monetary policy decision-making and to manage market expectations. Forecasts are thus tools rather than objectives, as the central bank’s role is to manage inflation risks rather than merely to produce accurate forecasts. It is thus important to recognize that the Bank of England’s organizational responsibilities shape its forecasting approach.  These responsibilities, which include transparency and accountability,  while not necessarily in conflict, influence how forecasts are generated and utilised, highlighting the complex role forecasts play within the Bank's broader policy framework.
    
This paper discusses three important themes highlighted in the review that warrant deeper consideration to understand the role of economic forecasting in the context of monetary policy. First, economic forecasting presents significant challenges, underlining the importance of judgment, the use of multiple forecasts rather than relying on a single projection, and the communication of uncertainty around central forecasts.  Second, monetary policy forecasts are conditional on a path of future policy rates. If disclosing the policymaker's expected path for future policy rates is not feasible, it is preferable to use model-consistent (and thus unconditional) forecasts rather than relying on potentially inappropriate assumptions. Third, forecast evaluation is an essential step of the forecasting process as it allows to identify directions for improvement. This evaluation should be robust to forecast instabilities and misspecification of the central bank’s loss function. Finally, a formal evaluation of the Bank of England's inflation forecasts shows that,  despite the large forecast errors in recent years, these were still relatively accurate,  as they significantly outperformed common time series benchmarks and their performance was comparable to that of professional forecasters.

The paper is organised as follows. Section~\ref{sec_forecasting} examines the challenges in economic forecasting and the implications for central bank forecasting practices. Section~\ref{sec_conditioning} discusses the role of assumptions about the future policy stance and Section~\ref{sec_evaluation} analyses forecast evaluation. Section~\ref{sec_BOE} presents an evaluation of the Bank of England's inflation forecasts, and Section~\ref{sec_conclusion} concludes.

\section{The imperfect science of economic forecasting}\label{sec_forecasting}

Despite advances in economic modeling, economic forecasting remains fraught with challenges. For this reason, economic forecasting is often considered as much an art as a science.

Forecasting models are often deliberately kept relatively simple to avoid excessive complexity, which can obscure the understanding of transmission mechanisms---an essential aspect of effective central bank communication. While more complex models might seem to offer greater accuracy, they often lack parsimony, leading to increased estimation errors and reduced forecast accuracy due to a higher number of parameters. However, any economic model can fail to provide accurate forecasts for three reasons.

First, the economy is a dynamic system characterized by evolving human behaviour and decision-making. Economic relationships change over time \cite{stock2002has,cogley2005drifts,primiceri2005time,cogley2010inflation}, and, to obtain accurate forecast,  it is crucial to recognize structural changes as they occur, which poses a significant challenge \cite{hendry2005forecasting,clements2006forecasting,castle2010forecasting,d2013macroeconomic,pettenuzzo2017forecasting}, see also \citeasnoun{Castle2024} for an application to UK inflation forecasting during 2021–24. Structural changes also complicate the forecast assessment, as ex ante rational forecasts can appear inefficient ex post \cite{timmermann2006forecast}.

Second, forecasting is further complicated by the occurrence of unanticipated shocks, such as natural disasters, geopolitical events, and financial crises. These shocks are, by definition, unpredictable and can cause significant deviations from expected economic paths, thereby reducing the accuracy of forecasts. For example, the COVID-19 pandemic and the subsequent global recession were unexpected events that dramatically disrupted economic activities worldwide. The rapid spread of the virus led to unprecedented government interventions, such as lockdowns and fiscal stimulus measures, which could not be anticipated and significantly impacted the accuracy of economic forecasts for this period.

Third, the availability and timeliness of data are also major constraints in economic forecasting. Macroeconomic data are often released with a delay, forcing policymakers to make decisions that will affect future outcomes without complete information about the current policy stance \cite{orphanides2005reliability}.
As a result, forecasting future developments requires first nowcasting the current economic conditions by incorporating any new information that has become available since the latest data release of the variables of interest \cite{croushore2006forecasting,giannone2008nowcasting}. This is a crucial step for constructing accurate forecasts, as the nowcast provides the origin for future projections \cite{sims2002role,wright2009forecasting}.
In addition,  data revisions pose a challenge for assessing forecast accuracy, as the forecasting performance depends on which vintage of data is used for forecast evaluation. In particular, for GDP growth, inaccurate real-time data and their subsequent revisions cause serious difficulties for forecast construction and evaluation \cite{boero2008evaluating}, while the real-time data flow is much less of an issue for inflation \cite{patton2011predictability}.

\subsection{Implications for central bank forecasting practices}\label{sec_implications}

The main implication of these challenges is that simple naive forecasts are often hard to consistently outperform, as they are robust and less prone to overfitting. In particular, there is often no single forecasting method that consistently outperforms naive predictions over extended evaluation periods \cite{banerjee2006there}. For this reason, central banks often use a combination of models, as forecast combinations produce better results on average than methods based on the ex-ante best individual forecasting model \cite{bates1969combination,clemen1989combining,stock2004combination,hendry2004pooling}. The advantage of forecast combinations stems from their ability  to harness diversification gains, making them more robust to model misspecification, instability, and estimation error \cite{timmermann2006forecast}.
Simple averages of forecasts from different forecasting models are difficult to outperform \cite{genre2013combining}, but, for inflation forecasting, there is  also some evidence that Bayesian model averaging can outperform simple averages \cite{wright2009forecasting,koop2012forecasting,groen2013real}.

To enhance their forecasting performance, central banks also routinely incorporate expert judgment to refine model predictions \cite{mcnees1990role,sims2002role,alessi2014central}. This process includes combining forecasts from different models and integrating information that models may not capture effectively, such as new real-time information, official announcements, and unexpected events. The practice is supported by the evidence that improving on survey forecasts is challenging in real-time  \cite{faust2009comparing,croushore2010evaluation}. For example, \citeasnoun{ang2007macro} find that surveys can forecast US inflation better than time series models and financial variables.  However, as noted in \citeasnoun{stekler2007future},  it is important to maintain records that explain the reasons for making expert adjustments, as they can provide ex-post insights into the rationale behind judgmental interventions.

The \citeasnoun{bernanke2024forecasting}  review documents that the Bank of England, like other central banks, employs a suite of models to analyze various aspects of the economy, and then it combines the model forecasts with ``substantial'' judgment and diverse information from external sources. While acknowledging that human judgment is a crucial element of real-world forecasting, the review cautions that ad hoc adjustments may obscure deeper issues within the forecasting framework or fail to reflect structural changes in the economy. This concern is well-founded, as relying too heavily on judgmental interventions without thorough documentation risks introducing biases and undermining the transparency of the forecasting process. To mitigate this risk, the review recommends maintaining detailed records of judgmental interventions. Such records enable retrospective analysis of how these adjustments may have contributed to forecast errors and provide the basis for systematic evaluation and improvement of the forecasting framework.

Finally, due to the uncertainties and risks associated with economic forecasting, a number of inflation-targeting central banks provide a measure of uncertainty to complement their point forecasts, and also collect survey respondents' beliefs about the probability distribution around the point forecast. Point forecasts can overlook the dispersion of possible outcomes and do not allow the analysis of the risks surrounding the central scenario. These risks are typically not symmetrically distributed, which can matter if the policymaker has an asymmetric loss function, for example, facing a higher loss if the realised inflation is above the projected level than if it is lower instead. Traditional approaches for estimating the full predictive distributions of the target variables require a large number of restrictive assumptions for tractability or heavily rely on judgments, such as, for example, the Bank of England fan charts \cite{britton1998inflation,clements2004evaluating}. Indeed, the \citeasnoun{bernanke2024forecasting} review notes that the Bank of England fan charts have ``weak'' conceptual foundations, convey limited useful information beyond what could be communicated through more direct ways, and attract little attention from the public. Consequently, the review recommends eliminating fan charts and replacing them with alternative scenarios to help the public better understand the rationale behind policy choices, including risk management considerations.

Although it is a valid question how the fan charts are estimated and to what extent they are determined by judgment, the proposal in the \citeasnoun{bernanke2024forecasting} review represents a significant shift. Scenario analysis and density forecasting are not substitutes, but rather complementary tools, particularly when risks are asymmetrically distributed. Scenario analysis allows to assess the implications of specific risks, while density forecasting offers a probabilistic view of the potential outcomes. Selecting which risks to highlight requires careful judgment, as the omitted scenarios may carry important implications for policymaking. In addition, as the number of risks increases, the proliferation of potential scenarios makes it challenging to decide which ones to select, complicating effective communication and potentially diluting key messages conveyed to the public. 

On the other hand, recent advancements in large Bayesian VAR modelling offer a valid alternative for inflation density forecasting \cite{groen2013real,carriero2016common,koop2013large,giannone2015prior}. Another approach that is gaining traction among central banks is to model directly the conditional quantiles of the predictive distribution of GDP \cite{adrian2019vulnerable,lloyd2024foreign} and of inflation \cite{manzan2013macroeconomic,lopez2024inflation}.
Therefore, while alternative scenarios can improve transparency and enhance the public's understanding of policy choices, density forecasting remains a valuable tool for capturing the full spectrum of risks. Combining these methods, rather than replacing one with the other, may provide the most effective way for communicating uncertainty and guiding monetary policy decisions.

\section{The conditioning policy path dilemma}\label{sec_conditioning}
Future outcomes of macroeconomic variables, such as real GDP growth and inflation, depend on the future course of monetary policy. This dependency makes the forecasting process for central banks particularly sensitive, as it requires the central bank to make assumptions about its own future policy stance.

A simple strategy is to base economic projections on the assumption of a constant policy rate--that is, assuming that the policy rate will remain unchanged from its current level. However, this assumption can be unrealistic during periods of macroeconomic distress, and, by not accounting for potential policy adjustments, the derived forecasts can be unreliable precisely when accurate predictions are most critical.

\begin{figure}[t!]
\caption{Bank Rate and Market Expectations}\label{fig_bank_rate}
	\begin{center}
	\includegraphics[trim=2cm 15cm 2.5cm 8cm, clip, scale=0.8]{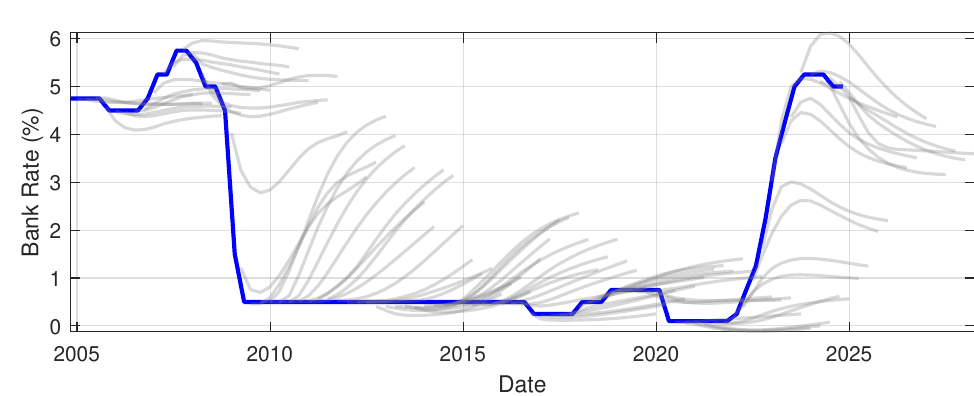}
\end{center}\clearpage
	{\footnotesize Note: Bank of England's bank rate and the 15-day averages of overnight forward rates from the commercial bank liability curve, adjusted for credit risk. Source: Bank of England.}
\end{figure}

An alternative strategy, employed by institutions such as the European Central Bank and the Bank of England, involves conditioning central bank forecasts on market expectations of future policy rates, thereby integrating market expectations of future policy changes into the projections. However, financial markets have often failed to predict central banks' actions. Evidence suggests that market expectations can be biased predictors of future short rates \cite{soderstrom2001predicting,schmeling2022monetary} and are more dispersed when inflation volatility is driven by supply shocks \cite{madeira2023origins}. Figure~\ref{fig_bank_rate} reports the  Bank of England's bank rate and the market expectations used as conditioning paths by the Bank of England  since November 2004. The figure reveals that market expectations are poor predictors of the actual bank rate, as for example in 2010-11 markets were expecting a monetary policy tightening that did not materialise, and also in the most recent sample, when markets have systematically been expecting a looser policy than the one implemented by the Bank of England. Relying on market expectations may thus distort the central bank's forecasts, as these expectations often do not accurately reflect the most likely future path of monetary policy or the true views of the policymaker. Also, market expectations might be inconsistent with alternative scenarios that focus on specific risks and that might require different policy adjustments. 

Despite these drawbacks, using constant rates or market-based expectations is appealing for central banks because it avoids the need for the central bank to make explicit assumptions about its own future policy stance. There is often the concern that these assumptions could be interpreted as a statement of the bank's intentions \cite{goodhart2009interest}, and indeed there is some empirical evidence that 
previously announced interest rate paths might influence current policy decisions \cite{mirkov2016announcements}. However, in an economy where private agents have imperfect information about the determination of monetary policy, central bank communication of interest rate projections can help shape financial market expectations and may improve the central bank's ability to achieve its stabilization objectives \cite{woodford2005central,rudebusch2008revealing}.  In particular, disclosing interest rate projections allows the central bank to communicate its private information about its preferences or beliefs \cite{bassetto2019forward}.

Since 2012, the Federal Reserve has published the interest rate projections of individual Federal Reserve Open Market Committee members in the so-called ``dot-plot".   The increased transparency introduced by publishing  Federal Reserve interest rate projections has reduced market uncertainty about the future path of US monetary policy \cite{el2016revisiting,guisinger2023reconsidering}, which is crucial in order to tackle market volatility. The Federal Reserve interest rate projections have also been shown to provide guidance about the long-run path of interest rates \cite{hillenbrand2021fed} and 
additional information to macroeconomic projections \cite{hofmann2022quantitative}.  

Despite this evidence, the \citeasnoun{bernanke2024forecasting}  review does not recommend adopting a dot-plot approach similar to that of the Federal Reserve. The review acknowledges the potential drawbacks of conditioning on market expectations, but suggests that disclosing individual or aggregate forecasts of the policy rate could be highly ``consequential'', leaving decisions on this issue to future deliberations. Instead, it recommends de-emphasizing the central forecast conditioned on the market rate path and placing more emphasis on publishing alternative scenarios, which can include an endogenous response of monetary policy to the assumed changes in the outlook. The review suggests that these responses could be generated by the staff in consultation with the Monetary Policy Committee, based on the historical behaviour of the Committee, policy rules or optimal policy calculations, allowing for a more realistic reflection of potential future policy adjustments in response to evolving economic conditions. However, while adopting an endogenous policy path for the alternative scenarios is a step in the right direction, the need to carefully select the assumptions about changes in the outlook makes it difficult to implement. Furthermore, there is a risk that market participants might focus on the alternative scenarios over the central forecast based on market expectations, potentially rendering the latter redundant.

An alternative way of addressing the dilemma of potentially inappropriate conditioning assumptions, without requiring the policy maker to disclose the likely future path of interest rates, is to fully delegate the responsibility for producing central bank forecasts to central bank staff.   The appropriate division of labour between the central bank's professional staff and the appointed policymakers is a key issue in monetary policymaking. The \citeasnoun{bernanke2024forecasting}  review documents that in peer central banks the involvement of policymakers in forecast construction is generally less extensive than at the Bank of England. Delegating the responsibility for producing central bank forecasts to central bank staff would free up the Monetary Policy Committee to focus on managing inflation risks and  would align better with the practices of the Bank of Canada, the Norges Bank, the Riksbank, and the Reserve Bank of New Zealand, all of which use endogenous policy paths.

\citeasnoun{ellison2012defense}, however, argue that policymakers’ forecasts often carry stronger strategic and policy content compared to staff forecasts, which are generally more focused on predicting future outcomes, and thus are more akin to market forecasts. \citeasnoun{guisinger2023reconsidering} further show that the Federal Reserve’s inflation forecasting advantage, as highlighted by \citeasnoun{romer2000federal}, stemmed from the market needing to predict both economic shocks and policy actions, while the Federal Reserve itself had greater certainty about its own intentions. On balance, while conditioning on the likely policy path is the most effective approach for obtaining accurate forecasts, delegating forecast responsibilities to staff provides a pragmatic compromise, allowing for the use of model-consistent paths without revealing policymakers’ views, thereby fostering a more open and transparent discussion about central bank forecast evaluation.

\section{Learning from errors}\label{sec_evaluation}

Accurate forecasts are the key to good decision-making, and determining their accuracy requires a formal forecast evaluation process. As emphasized in the \citeasnoun{bernanke2024forecasting}  review, systematically assessing forecast performance and understanding the underlying causes of forecast errors are essential steps for enhancing the overall quality and reliability of economic forecasting.

Central bank forecasts are typically evaluated using various methods. While summary statistics of forecast errors, as reported in the Bank of England's Monetary Policy Report and in the \citeasnoun{bernanke2024forecasting} review, provide an initial overview of forecast accuracy, these descriptive measures alone are insufficient for formal forecast evaluation, which instead can be implemented  through forecast evaluation tests. Specifically, forecast rationality tests, such as \citeasnoun{mincer1969evaluation} regressions and their generalizations in \citeasnoun{ericsson2017biased}, can identify potential departures from rationality and point to directions for forecast improvement. In addition, forecast comparison tests, including the \citeasnoun{diebold1995comparing} test of equal predictive accuracy and the  forecast encompassing test \cite{chong1986econometric,harvey1997testing}, formally compare the performance of different forecasting methods. However, two important issues that should be considered in the forecast evaluation process are forecast instabilities and loss function dependence.

As noted in the \citeasnoun{bernanke2024forecasting} review, due to structural changes in the economy, the most accurate forecasting model may vary over time. This implies that local measures of models' forecasting performance are more appropriate than traditional full-sample average measures \cite{rossi2021forecasting}. Evaluating forecasts on subsamples of the original data can reveal how model performance evolves, but forecast evaluation tests can suffer from small sample biases when the number of out-of-sample observations is relatively small \cite{clark2013advances}. Fixed-smoothing asymptotics can address these small-sample size distortions and provide robust test results even in small samples \cite{coroneo2020comparing,harvey2017forecast,coroneo2024survey}. In addition, fluctuation tests \cite{giacomini2010forecast,rossi2016forecast} allow to repeatedly test the forecasting performance in  rolling windows  to assess the relative local forecasting performance.

Finally, the policymakers' optimal forecasts is determined by their preferences and their aversion to different types of forecasting errors. This implies that the central bank's loss function has a crucial role in the forecast evaluation process. Most of the literature has focussed on squared error loss, partly because traditional tests for rationality and encompassing rely on this assumption. More recently, due to the large forecast error due to the Covid-19 pandemic and the subsequent spike in inflation, the absolute error loss has gained popularity \cite{alessi2014central,Kanngiesser_Willems_2024boe}. However, there is growing evidence that central banks exhibit asymmetric preferences \cite{robert2003optimal,ruge2003inflation,capistran2008bias}. As a result, under standard tests, central bank forecasts can appear biased and inefficient, since the standard properties of optimal forecasts can be invalid under asymmetric loss \cite{elliott2008biases}. In such cases, robust testing methodologies should be used instead, as for example rationality test methodologies that are valid for flexible or unknown losses  \cite{elliott2005estimation,patton2007testing}, and robust forecast comparison tests that do not depend on a specific loss function \cite{jin2017robust,corradi2023robust}.

\section{A formal evaluation of the Bank of England's inflation forecasts}\label{sec_BOE}
\begin{figure}[tbp!]
\caption{UK CPI inflation forecasts}\label{fig_forecasts}
	\begin{center}
	\includegraphics[trim=2cm 6cm 2.5cm 7cm, clip, scale=0.8]{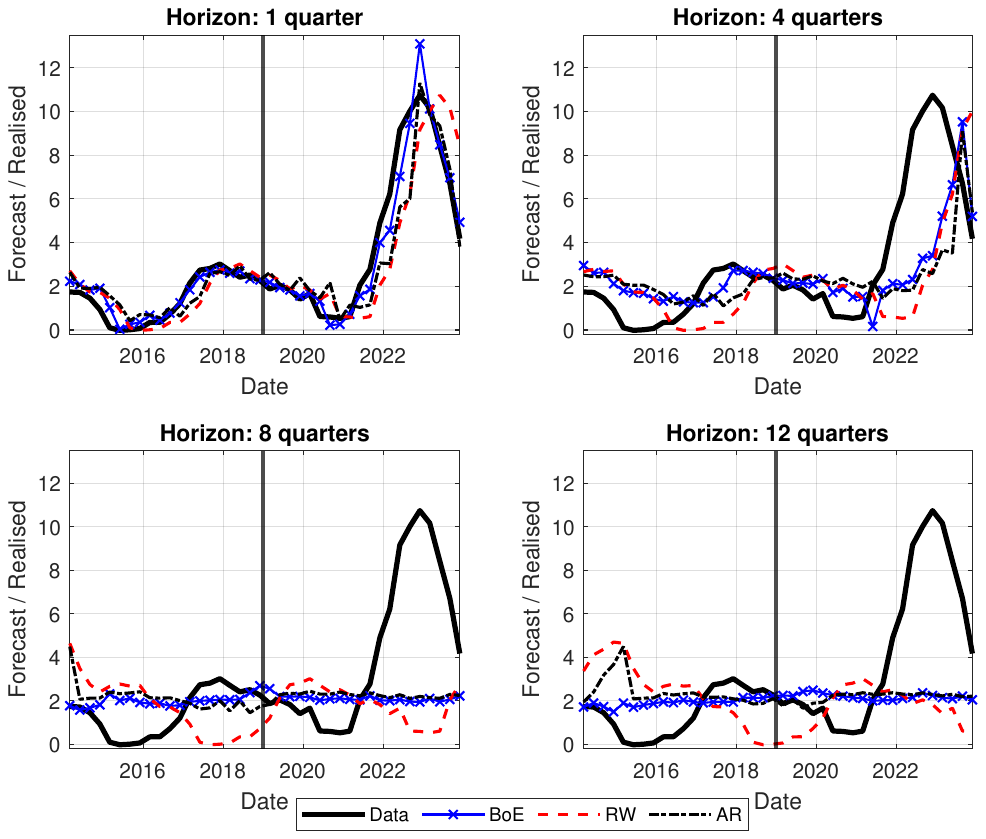}
\end{center}\clearpage
	{\footnotesize Note: Bank of England (BoE) inflation forecasts at forecasting horizons of 1, 4, 8 and 12 quarters, along with realized CPI inflation and the forecasts from the random walk (RW) and the autoregressive (AR) benchmarks. Quarterly observations from 2014.Q1 to 2023.Q4. The vertical line indicates the mid-sample point 2018.Q4 and delimits the two sub-samples.}
\end{figure}

This section contains a formal evaluation of the performance of the Bank of England's inflation forecasts over recent years. Figure~\ref{fig_forecasts} reports the Bank of England modal inflation forecasts for horizons one to twelve quarters ahead for the period 2014.Q1 to 2023.Q4 along with the realised CPI inflation, and the forecasts from two simple benchmarks.\footnote{I thank Tim Willems for sharing the Bank of England forecasts used in \citeasnoun{Kanngiesser_Willems_2024boe}}The first one is the random walk, for which the forecast for any horizon is given by the latest quarterly inflation available before the Monetary Policy Report release. The second one is the autoregressive model with coefficients estimated on a rolling window of 60 observations and lag order selected using the  Akaike information criterion (with a maximum lag of four).\footnote{This analysis focuses on point forecasts, however, the forecast evaluation approach presented here can also be used to compare predictive distributions, as showed in \citeasnoun{coroneo2024survey}.} 

The figure shows that UK CPI inflation was low and stable in the first part of the sample, while in the second part, we can see the inflation surge of 2022. The Bank of England short-horizon forecasts are close to the benchmarks but distinct, this is particularly evident for 2022.Q4 when the Bank of England overshot its one quarter ahead forecast with a prediction of  13.1\%, while the realised inflation  was  10.76\%. At longer horizons, the Bank of England forecasts are more closely anchored to the 2\% inflation target, while the two benchmarks display more variability.

\begin{figure}[tbp!]
\caption{Forecast errors}\label{fig_e}
	\begin{center}
	\includegraphics[trim=2cm 6cm 2.5cm 7cm, clip, scale=0.8]{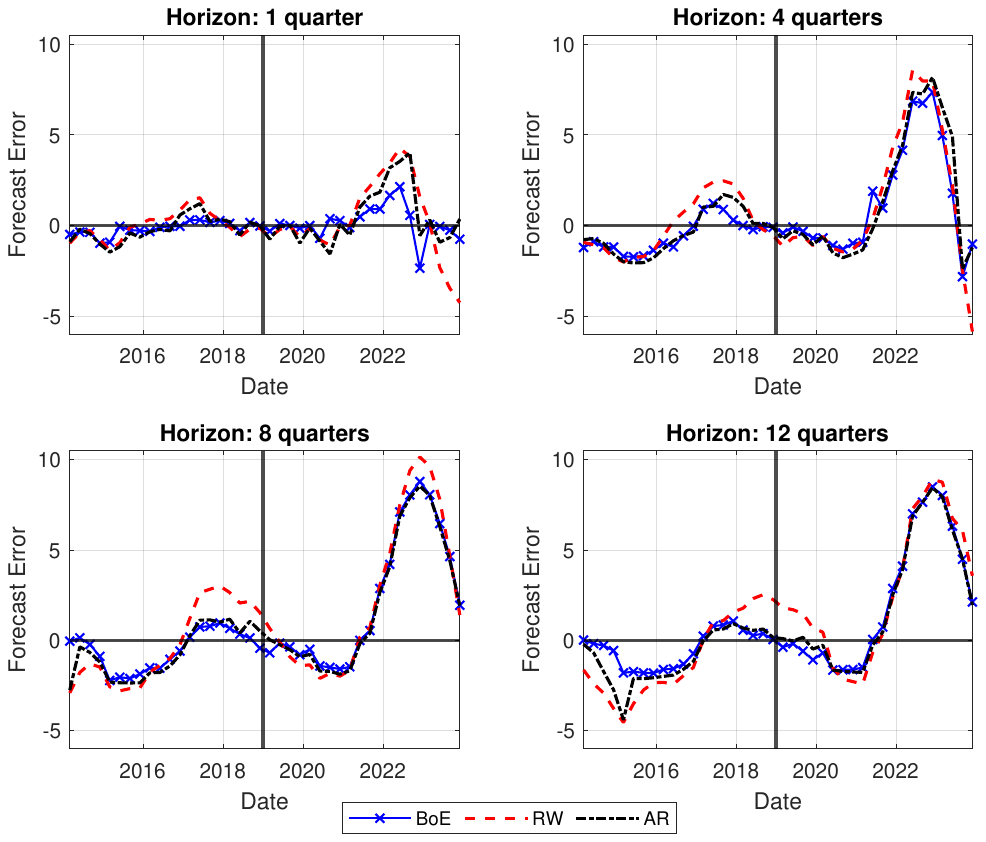}
\end{center}\clearpage
		{\footnotesize Note: Bank of England (BoE) inflation forecast errors at forecasting horizons of 1, 4, 8 and 12 quarters, along with the forecast errors from the random walk (RW) and the autoregressive (AR) benchmarks.  Forecast errors are defined as the realized value minus the forecast. Quarterly observations from 2014.Q1 to 2023.Q4. The vertical line indicates the mid-sample point 2018.Q4 and delimits the two sub-samples. }
\end{figure}

\begin{singlespace}
    \begin{table}[htbp]
 \center
  \textbf{\caption{Summary Statistics of Forecast Errors\label{tab_e}}}\thispagestyle{empty}
\medskip
\resizebox{0.65\textheight}{!}{
  \centering
    \begin{tabular}{r|r|*{9}{C}}%
    \toprule
     \multicolumn{11}{c}{\textbf{First Sub-sample: 2014.Q1 to 2018.Q4 }} \\
     \midrule
     &   h & Mean & Med & MAE & MdAE & Std & Max & Min &  AC1 & AC4 \\
    \midrule
        \multicolumn{1}{c|}{\multirow{6}[0]{*}{BoE}}
        & 0 & -0.06 & -0.04 & \textbf{0.10} & \textbf{0.08} & \textbf{0.12} & \textbf{0.14} & \textbf{-0.35} &  0.40 & -0.16 \\
        & 1 & -0.16 & \textbf{-0.07} & \textbf{0.29} & \textbf{0.28} & \textbf{0.36} & \textbf{0.31} & \textbf{-0.97} &  \textbf{0.63} & 0.44 \\
        & 2 & -0.30 & \textbf{-0.11} & \textbf{0.49} & \textbf{0.40} & \textbf{0.60} & \textbf{0.53} & -1.80 &  \textbf{0.69} & 0.51 \\
        & 4 & -0.54 & -0.73 & \textbf{0.87} & \textbf{0.93} & \textbf{0.91} & \textbf{1.22} & \textbf{-1.73} &  \textbf{0.90} & 0.32 \\
        & 8 & -0.53 & \textbf{-0.33} & \textbf{0.92} & \textbf{0.76} & \textbf{1.06} & \textbf{0.96} & \textbf{-2.22} &  0.89 & \textbf{0.12} \\
        &12 & \textbf{-0.46} & \textbf{-0.23} & \textbf{0.88} & \textbf{0.77} & \textbf{1.01} & 1.07 & \textbf{-1.81} &  0.90 & \textbf{0.17} \\
\midrule
        \multicolumn{1}{c|}{\multirow{6}[0]{*}{RW}}
        & 0 &  \textbf{0.01} & \textbf{0.02} & 0.31 & 0.27 & 0.41 & 0.93 & -0.83 & 0.63 & \textbf{-0.04} \\
        & 1 &  \textbf{0.00} & -0.10 & 0.59 & 0.38 & 0.76 & 1.53 & -1.36 &  0.81 & \textbf{0.10} \\
        & 2 & \textbf{-0.01} & -0.22 & 0.87 & 0.83 & 1.06 & 2.02 & \textbf{-1.62} &  0.88 & \textbf{0.14} \\
        & 4 &  \textbf{0.00} & \textbf{-0.39} & 1.32 & 1.31 & 1.54 & 2.47 & -2.00 &  0.94 & \textbf{0.31} \\
        & 8 & \textbf{-0.22} & -1.15 & 2.03 & 2.12 & 2.21 & 3.01 & -2.91 &  0.95 & 0.62 \\
        &12 & -0.97 & -1.80 & 2.23 & 2.32 & 2.30 & 2.53 & -4.55 &  0.96 & 0.73 \\
\midrule
    \multicolumn{1}{c|}{\multirow{6}[0]{*}{AR}}
        & 0 & -0.08 & -0.06 & 0.27 & 0.26 & 0.34 & 0.68 & -0.81 &  \textbf{0.27} & -0.08 \\
        & 1 & -0.20 & -0.25 & 0.57 & 0.43 & 0.68 & 1.21 & -1.47 &  0.71 & 0.22 \\
        & 2 & -0.30 & -0.43 & 0.78 & 0.71 & 0.94 & 1.43 & -1.90 &  0.85 & 0.25 \\
        & 4 & -0.42 & -0.64 & 1.09 & 1.04 & 1.22 & 1.71 & -2.06 &  0.91 & 0.35 \\
        & 8 & -0.67 & -0.74 & 1.34 & 1.16 & 1.42 & 1.17 & -2.76 &  \textbf{0.85} & 0.47 \\
        &12 & -0.93 & -0.92 & 1.35 & 1.04 & 1.46 & \textbf{0.93} & -4.40 &  \textbf{0.84} & 0.33 \\
    \midrule
    \multicolumn{11}{c}{}\\
     \midrule
     \multicolumn{11}{c}{\textbf{Second Sub-sample:
     2019.Q1 to 2023.Q4}} \\
    \midrule
         &   h & Mean & Med & MAE & MdAE & Std & Max & Min &  AC1 & AC4 \\
        \midrule
        \multicolumn{1}{c|}{\multirow{6}[0]{*}{BoE}}
        & 0 & \textbf{0.08} & 0.06 & \textbf{0.21} & \textbf{0.14} & \textbf{0.26} & \textbf{0.58} & \textbf{-0.46} &  0.18 & \textbf{0.05} \\
        & 1 & \textbf{0.14} & 0.03 & \textbf{0.61} & \textbf{0.34} & \textbf{0.92} & \textbf{2.14} & -2.35 &  \textbf{0.35} & \textbf{0.01} \\
        & 2 & \textbf{0.44} & \textbf{0.10} & \textbf{1.13} & \textbf{0.57} & \textbf{1.60} & \textbf{4.37} & -2.39 &  \textbf{0.66} & -0.23 \\
        & 4 & \textbf{1.36} &\textbf{-0.20} & \textbf{2.39} & \textbf{1.20} & \textbf{3.09} & \textbf{7.35} & -2.82 &  \textbf{0.82} & \textbf{-0.03} \\
        & 8 & 2.21 & 0.27 & 3.06 & \textbf{1.54} & 3.73 & 8.80 & \textbf{-1.59} &  \textbf{0.93} & 0.42 \\
        &12 & \textbf{2.13} & 0.40 & 3.06 & \textbf{1.63} & 3.67 & 8.50 & \textbf{-1.64} &  0.94 & 0.45 \\
\midrule
        \multicolumn{1}{c|}{\multirow{6}[0]{*}{RW}}
        & 0 & 0.10 & \textbf{0.03} & 0.97 & 0.72 & 1.32 & 2.95 & -2.53 &  0.67 & -0.27 \\
        & 1 & 0.31 &-0.06 & 1.67 & 1.29 & 2.28 & 4.26 & -4.25 &  0.89 & -0.31 \\
        & 2 & 0.61 &-0.41 & 2.31 & 1.34 & 3.11 & 6.40 & -6.00 &  0.87 & -0.30 \\
        & 4 & 1.37 &-0.59 & 3.09 & 1.80 & 3.98 & 8.56 & -5.84 &  0.87 & -0.11 \\
        & 8 & 2.44 & 0.69 & 3.57 & 1.90 & 4.35 & 10.15 & -2.10 &  \textbf{0.93} & \textbf{0.39} \\
        &12 & 2.66 & 1.75 & 3.56 & 2.34 & 3.81 & 8.92 & -2.41 &  \textbf{0.93} & 0.34 \\
\midrule
    \multicolumn{1}{c|}{\multirow{6}[0]{*}{AR}}
        & 0 & 0.17 & -0.07 & 0.62 & 0.35 & 0.91 & 2.71 & -1.37 & \textbf{-0.10} & -0.15 \\
        & 1 & 0.45 & \textbf{-0.02} & 1.15 & 0.81 & 1.60 & 4.01 & \textbf{-1.57} &  0.63 & -0.17 \\
        & 2 & 0.75 & -0.37 & 1.79 & 1.06 & 2.49 & 6.26 & \textbf{-1.68} &  0.72 & \textbf{-0.12} \\
        & 4 & 1.49 & -0.39 & 2.80 & 1.54 & 3.59 & 8.17 & \textbf{-2.39} &  0.83 & 0.09 \\
        & 8 & \textbf{2.08} & \textbf{0.22} & \textbf{3.04} & 1.79 & \textbf{3.71} & \textbf{8.55} & -1.87 &  0.94 & 0.42 \\
        & 12 & 2.14 & \textbf{0.32} & \textbf{2.93} & 1.76 & \textbf{3.59} & \textbf{8.45} & -1.77 &  \textbf{0.93} & 0.39 \\
    \bottomrule
    \multicolumn{11}{p{15cm}}{\footnotesize {Note: The table reports summary statistics of forecast errors for the Bank of England (BoE) inflation forecasts, along with the ones for the random walk (RW) and the autoregressive (AR) benchmarks. The table reports mean, median, MAE (Mean Absolute Error), MdAE (Median Absolute Error), standard deviation (std), skewness (skew), and autocorrelation coefficients of order 1 and 4 (AC1 and AC4). Forecast errors are defined as the realized value minus the forecast. The (absolute) lowest number for each measure and horizon is highlighted in bold.}}
    \end{tabular}}
\end{table}%
\end{singlespace}
\clearpage

Forecast errors, defined as the realized value minus the forecast, are reported in Figure~\ref{fig_e}. The figure indicates a change in the volatility of the forecast errors between the first and the second halves of the sample. Accordingly, Table~\ref{tab_e} reports summary statistics of the forecast errors separately for the two half sub-samples, where the first sub-sample is from 2014.Q1 to 2018.Q4, and the second sub-sample from 2019.Q1  to 2023.Q4. In the first sub-sample, all the forecasts exhibit negative mean and median forecast errors,  due to the fact that all the forecasts over-predicted inflation. In the second subsample, instead, all forecasts have a positive mean error, indicating under-prediction. In addition, in the second subsample, there is a greater difference between mean and median error, as the inflation surge of 2022 was accompanied by extreme forecast errors, indeed the maximum forecast errors in the second sub-sample reach extreme values above 8\% for four quarters ahead forecasts and beyond. The larger mean errors, mean absolute errors and standard deviations also reveal the challenges with forecasting inflation in the second sub-sample. The comparison of the mean absolute errors, median absolute errors and standard deviations suggests that in both sub-samples the Bank of England inflation forecasts are generally more accurate than the two benchmarks,  in line with \cite{Kanngiesser_Willems_2024boe}, who also ran a horse race of the Bank of England inflation forecasts against the random walk and autoregressive benchmarks, arriving at similar conclusions. The table also indicates that the autoregressive benchmark is more accurate in the second subsample for 8 and 12 quarters ahead forecasts, in line with the earlier finding in \cite{kapetanios2008forecast} that the autoregressive benchmark is hard to beat.

To formally compare the predictive accuracy of the Bank of England forecasts with the two benchmarks, I use the \citeasnoun{diebold1995comparing} test of equal predictive accuracy. This approach adopts a model-free perspective to compare two competing forecasts, imposing assumptions only on
the forecast error loss differential.\footnote{An alternative approach that allows for multiple comparison of forecasting models is the model confidence set of \citeasnoun{hansen2011model}. This approach allows to select  a set of models (not necessarily one) that it will contain the best model with a given level of confidence.}
For a given loss function, the null hypothesis of the \citeasnoun{diebold1995comparing} test is that the expected loss differential of two competing forecasts is zero. Given that the test relies on the choice of a loss function, I consider both the popular quadratic and absolute value loss functions, as well as the linear exponential (linex) loss function, which is an asymmetric loss function that allows to give larger penalty to over-prediction or under-prediction \cite{christoffersen1997optimal}. In particular, I consider two parametrisations of the linex function $\exp(\alpha e_t)-\alpha e_t-1$, where $e_t$ is the forecast error, by using two values for $\alpha$, namely, 0.5 and -0.5, the former penalises more under-predictions and the latter over-predictions. 

As indicated by Figure~\ref{fig_e}, the mid- to long-horizon forecast errors exhibit some degree of autocorrelation. Because strong autocorrelation can lead to size distortion or low power in the \citeasnoun{diebold1995comparing} test \cite{coroneo2024dependence}, I further investigate this issue by reporting the autocorrelation coefficients in Table~\ref{tab_e}. These results show that although the first-order autocorrelation is substantial, it decays quickly, as evidenced by the fourth-order autocorrelation, implying that the forecast errors are only weakly dependent. Thus, in this application, the \citeasnoun{diebold1995comparing} test should not be adversely affected by the issues identified in \citeasnoun{coroneo2024dependence}. Nonetheless, to avoid the small sample size bias of the \citeasnoun{diebold1995comparing} test, I use critical values derived under fixed-$b$ asymptotics as proposed in \citeasnoun{coroneo2020comparing}. With this alternative asymptotics,  the test of equal predictive accuracy has a nonstandard limit distribution that depends on the bandwidth to sample size ratio and on the kernel used to estimate the long-run variance, see \citeasnoun{kiefer2005new} for critical values. 

Given the differences highlighted in the previous discussion, I perform the forecast evaluation on both the full sample (from 2014.Q1 to 2023.Q4) and also separately on two equally sized sub-samples. This means that for each evaluation sub-sample, inference is based on just 20 observations, exacerbating the small sample size bias of standard asymptotics and making fixed-smoothing crucial to obtain correctly sized tests \cite{coroneo2023testing}. In particular, I estimate the long-run variance of the loss differential using the Bartlett kernel with bandwidth $\lfloor T^{1/2}\rfloor$, which is equal to 6 for the full sample and 4 for the two sub-samples. The 10\% and the 5\% critical values using fixed-$b$ asymptotics are then $\pm 2.09$ and $\pm 2.57$, instead of  $\pm 1.65$ and $\pm 1.96$ under standard asymptotics.

\begin{figure}[tbph!]
\caption{Forecast evaluation}\label{fig_test}
	\begin{center}
	\includegraphics[trim=2cm 6cm 2cm 7cm, clip, scale=0.85]{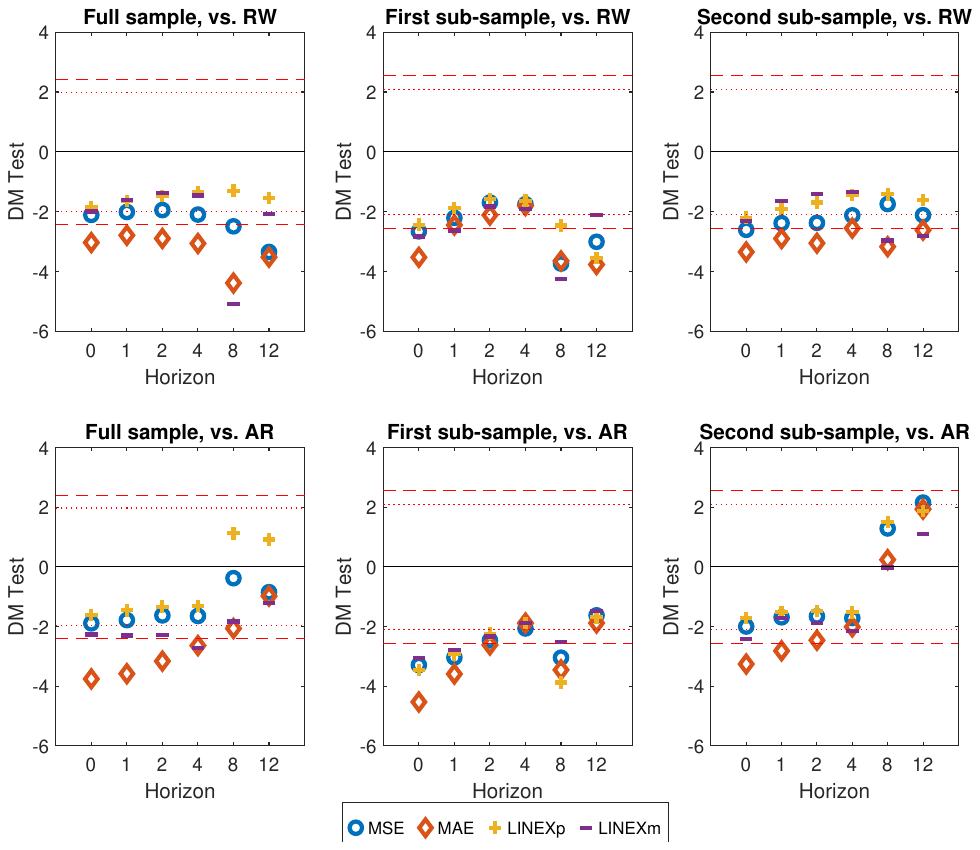}
\end{center}\clearpage
	{\footnotesize Note: \citeasnoun{diebold1995comparing} test statistic for the null of equal predictive accuracy of the Bank of England and the random walk (first row) or the autoregressive (second row) benchmarks. A negative value of the test statistic indicates a lower loss for the Bank of England, i.e. better performance with respect to the benchmark. Different loss functions are reported with different markers: a circle  refers to a quadratic loss function, a diamond to the absolute loss function, a plus to the linex with $\alpha=0.5$ and a minus to the linex with $\alpha=-0.5$. The horizontal axis denote the forecasting horizons (in quarters). The dotted and dashed red horizontal lines denote the 5\%, and 10\% significance levels using fixed-$b$ asymptotics as in \citeasnoun{coroneo2020comparing}. The full sample is from 2014.Q1 to 2023.Q4, the first sub-sample from 2014.Q1 to 2018.Q4 and the second sub-sample from 2019.Q4 to 2023.Q4.}
\end{figure}

Figure~\ref{fig_test} reports the test statistic for the null of equal predictive accuracy of the Bank of England and the random walk (first row) or the autoregressive (second row) benchmarks. A negative value of the test statistic indicates a lower loss for the Bank of England, i.e. better performance with respect to the considered benchmark. Different loss functions are reported with different markers, and the horizontal lines indicate the 5\% and the 10\% critical values using fixed-$b$ asymptotics.

Looking at the figure, we can draw two main results. First, the test statistics are negative for almost all samples, forecasting horizons, loss functions, and benchmarks, suggesting that the Bank of England’s inflation forecasts generally resulted in lower losses compared to the benchmarks. However, this outperformance is significant for all loss functions only for forecasts made for the current quarter and eight quarters ahead in the first sub-sample, while in the second sub-sample the forecast evaluation test statistics crucially depend on the choice of the loss function. Second, the null hypothesis of equal predictive accuracy is rejected more often when the absolute loss function is used. This is particularly evident in the second sub-sample, where, using the absolute loss, the Bank of England's inflation forecasts significantly outperform the random walk for all forecasting horizons and the autoregressive benchmark up to four quarters ahead.

\begin{figure}[tbp!]
\caption{Comparison with Survey Forecasts}\label{fig_survey}
	\begin{center}
	\includegraphics[trim=2cm 6.5cm 2.5cm 7cm, clip, scale=0.8]{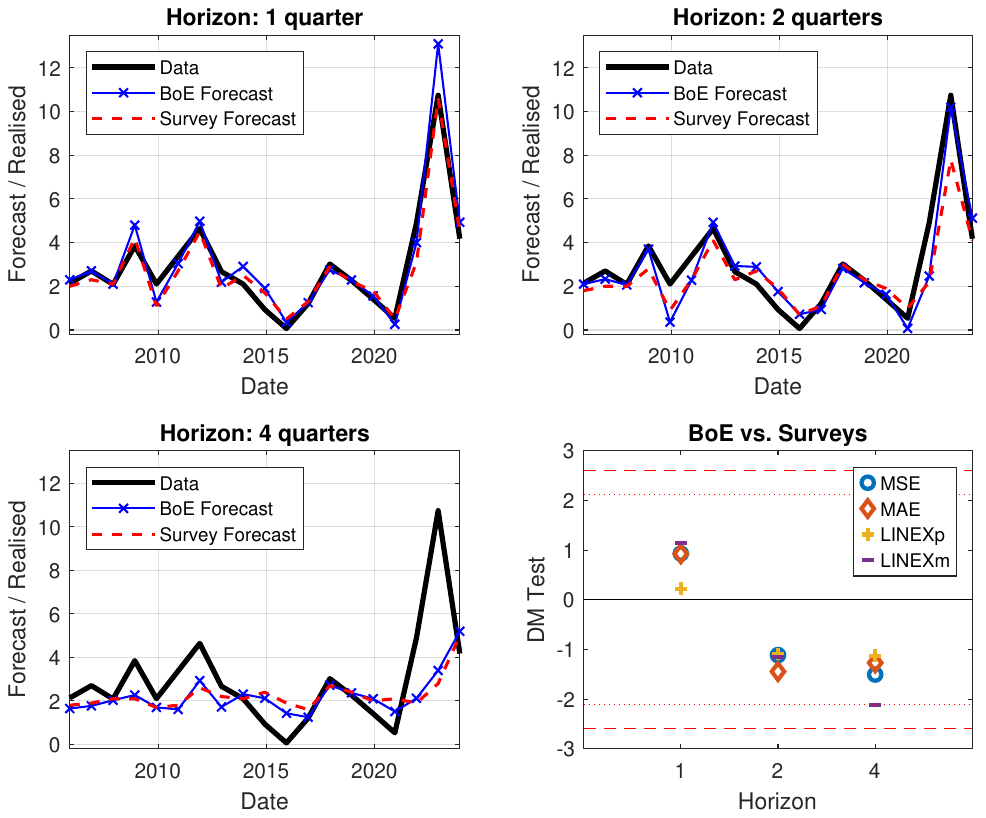}
\end{center}\clearpage
	{\footnotesize Note: Bank of England (BoE) inflation forecasts versus the survey forecasts published by the UK HM Treasury. The first three subplots report the forecasted and realised CPI inflation at horizons of 1, 2 and 4 quarters. The bottom left plot reports the 
    \cite{diebold1995comparing} test statistic for the null of equal predictive accuracy of the Bank of England and the survey forecasts. A negative value of the test statistic indicates a lower loss for the Bank of England, i.e. better performance with respect to the surveys. Different loss functions are reported with different markers: a circle  refers to a quadratic loss function, a diamond to the absolute loss function, a plus to the linex with $\alpha=0.5$ and a minus to the linex with $\alpha=-0.5$. The horizontal axis denote the forecasting horizons (in quarters). The dotted and dashed red horizontal lines denote the 5\%, and 10\% significance levels using fixed-$b$ asymptotics as in \citeasnoun{coroneo2020comparing}. The sample is from 2004 to 2023.}
\end{figure}

\citeasnoun{groen2009real} attribute the outperformance of the Bank of England inflation forecasts with respect to statistical benchmarks in the period 1997 to 2006 to the significant role of expert judgment in their forecasting process. To assess whether this result holds also in the most recent sample, I  compare the Bank of England inflation forecasts with a more challenging benchmark—survey forecasts. In particular, I consider the survey forecast published by the UK HM Treasury within the ``forecasts for the UK economy''.\footnote{The forecasts for the UK economy data is available at \url{https://www.gov.uk/government/collections/data-forecasts}.}  The panel is composed of about 30 professional forecasters (both City and non-City) that provide forecasts for a set of macroeconomic indicators for the current and the following year approximately by the 15th of each month. In particular, CPI inflation forecasts refer to CPI annual percentage changes for the current year Q4 and the following year Q4. To align the fixed-date survey forecasts (that refer to the end of year) to the Bank of England forecasts (that instead have fixed forecasting horizons), I consider  the Q4 forecast
made in August of the same year as the one quarter ahead forecast, the Q4 forecast made in May of the same year as the two quarters ahead forecast, and the Q4 forecast made in November of the previous year as the four quarters ahead forecast.

Figure~\ref{fig_survey} reports the survey forecasts for the 1, 2 and 4-quarter ahead horizons, along with the Bank of England forecasts and the realised inflation for 2004.Q4 to 2023.Q4.\footnote{Notice that, in this case, due the forecast horizon alignment, we have only one survey forecast per year, namely the one for Q4, therefore the sample size is 20 observations.}  The figure indicates that the survey forecasts are closely aligned with the Bank of England forecasts, in contrast with the two simple benchmarks in Figure~\ref{fig_forecasts} that displayed distinct dynamics. Indeed, the \citeasnoun{bernanke2024forecasting} review also notes that the forecast errors made by the Bank of England and those made by external forecasters are barely distinguishable. This finding aligns with the earlier observation in \citeasnoun{fildes2002state} that, despite adhering to different economic theories and using different methodologies, forecasters make similar mistakes, highlighting the significant role of judgment in the forecasting process. However, the alignment between the Bank of England and survey forecasts may also reflect their shared reliance on the same conditioning assumption for the policy path, i.e. market expectations.

The last panel of Figure~\ref{fig_survey} reports the test statistic for the null of equal predictive accuracy of the Bank of England and the survey forecasts. A negative value of the test statistic indicates a lower loss for the Bank of England, i.e. better performance with respect to the surveys. As in Figure~\ref{fig_test}, different loss functions are reported with different markers, and the horizontal lines indicate the 5\% and the 10\% critical values using fixed-$b$ asymptotics as in \citeasnoun{coroneo2020comparing}. The figure shows that the null of equal predictive accuracy is not rejected for any forecast horizon or loss function considered, with one notable exception: the four quarter ahead inflation forecast. At this horizon, the Bank of England significantly outperforms professional forecasters when using the asymmetric loss function that penalizes overpredictions more heavily, suggesting that professional forecasters tended to overpredict inflation at this horizon more than the Bank of England.

Taken together, these results highlight that, despite the larger forecast errors observed in 2022, the Bank of England’s inflation forecasts remained relatively accurate compared to both standard time series benchmarks and survey forecasts.

\section{Conclusion}\label{sec_conclusion}

Forecasting is a challenging yet crucial component of modern monetary policy. However, while accurate forecasting is important, the primary role of a central bank is to manage inflation risks, and, thus, central bank forecasts should not be judged solely on their accuracy.

The current disenchantment towards central bank forecasting brings us back to \citeasnoun{zarnowitz1991has} observation: ``The difficult question is how much of it is due to unacceptably poor performance and how much to unrealistically high prior expectations. My argument is that the latter is a major factor.” This suggests that the challenges faced in forecasting might be partly due to overly high expectations rather than solely poor performance. Indeed, the \citeasnoun{bernanke2024forecasting} review notes that ``given the unique circumstances of recent years, unusually large forecasting errors by the Bank [of England] during that period were probably inevitable''. 

Moving forward, it’s  crucial for the Bank of England to draw what lessons it can from the experience, but also to set realistic expectations for central bank forecast accuracy and to appreciate the broader role of central bank's forecasts in managing inflation risk.

\end{doublespace}

\addcontentsline{toc}{section}{References}
\bibliographystyle{cje}
\bibliography{ref}

\end{document}